\documentclass[onecolumn]{aa}
\usepackage[varg]{txfonts}
\usepackage{amsmath}
\usepackage{amsfonts}
\usepackage{amssymb}
\usepackage{textcomp}
\usepackage{mathrsfs}
\usepackage{bm}
\usepackage{hyperref}
\usepackage{color}

\usepackage{graphicx}
\usepackage{lmodern}
\usepackage{subfig}

\definecolor{CiteColor}{rgb}{0,0,0.35}
\definecolor{URLColor}{rgb}{0,0,0.35}
\hypersetup{colorlinks,citecolor=CiteColor,urlcolor=URLColor}

\newcommand{\beq}{\begin{equation}}
\newcommand{\eeq}{\end{equation}}
\newcommand{\dd}{\mathrm{d}}

\allowdisplaybreaks

\begin{document}

\title{Detecting the General Relativistic orbital precession\\ of the exoplanet HD\,80606b}

\author{Luc Blanchet\thanks{luc.blanchet@iap.fr} \and Guillaume H\'ebrard\thanks{hebrard@iap.fr} \and Fran\c{c}ois Larrouturou\thanks{francois.larrouturou@iap.fr}}

\institute{Institut d'Astrophysique de Paris,\\ UMR 7095, CNRS, Sorbonne Universit{\'e},\\ 98\textsuperscript{bis} boulevard Arago, 75014 Paris, France}

\abstract{We investigate the relativistic effects in the orbital motion of the exoplanet HD\,80606b with a high eccentricity of $e\simeq 0.93$. We propose a method to detect these effects (notably the orbital precession) based on measuring the successive eclipse and transit times of the exoplanet. In the case of HD\,80606b, we find that in ten years (after approximately 33 periods) the instants of transits and eclipses are delayed with respect to the Newtonian prediction by about three minutes due to relativistic effects. These effects can be detected by comparing at different epochs the time difference between a transit and the preceding eclipse, and should be measurable by comparing events already observed on HD\,80606 in 2010 with the \textit{Spitzer} satellite together with those to be observed in the future with the \textit{James Webb Space Telescope}.}


\maketitle

%%%%%%%%%%%%%%%%%%%%%%%
\section{Introduction} 
\label{sec:intro}

Whereas more than 4000 exoplanets have been discovered so far, the planet orbiting the G5 star HD\,80606 remains a remarkable and unique case. Its eccentricity is extremely high: $e \simeq 0.93$. HD\,20782b is the only planet reported to have a higher eccentricity~\citep{HD80606b_OT09} but its high $e$-value stands on one single measurement and has not been confirmed up to now. The high eccentricity of HD\,80606b was well established  as soon as the planet was discovered through radial velocity measurements~\citep{HD80606b_N01} and has been largely confirmed by subsequent observations. Today the most accurate system parameters are those determined by~\citet{HD80606b_H11}, who provide $e=0.9330 \pm 0.0005$.

The long orbital period of the planet ($P=111.4367 \pm 0.0004$~days) implied a low probability of the orbital plan to be aligned with the line of sight to the Earth. Still, HD\,80606b was successively discovered to pass behind its parent star (planetary eclipses) by~\citet{HD80606b_LDL09}, then in front of it (planetary transits) by~\citet{HD80606b_M09} \citep[see also][]{HD80606b_GM09,HD80606b_F09,HD80606b_W09,HD80606b_Hi10}. That configuration allows the planetary radius to be measured ($R_p = 0.981 \pm 0.023\,R_\mathrm{Jup}$); the inclination of the orbit is also known ($i = 89.269^{\circ} \pm 0.018^{\circ}$), which provides the mass $M_p = 4.08 \pm 0.14\,M_\mathrm{Jup}$ from the sky-projected mass derived with radial velocities.
The obliquity of the system could also be measured from the Rossiter-McLaughlin anomaly observed during transits, which revealed that the orbit of HD\,80606b is prograde but inclined~\citep{HD80606b_M09,HD80606b_P09,HD80606b_W09}; \citet{HD80606b_H11} reported an obliquity of $\lambda = 42^{\circ} \pm 8^{\circ}$. Such a peculiar orbit may be explained by the influence of the companion star HD\,80607, located 1200\,AU further away, through a Kozai-Lidov mechanism~\citep{HD80606b_P09,HD80606b_C11}. 

The particularly high eccentricity of the planetary orbit raises the question of the feasibility of detecting the general relativistic (GR) precession of its periastron.
Indeed it should be enhanced to
\begin{equation}\label{DeltaGR}
\Delta_\text{GR} = \frac{6\pi G M}{a c^2 (1-e^2)} \simeq
215~\text{arcsec}/\text{century}\,,
\end{equation} 
as compared to $43\,\text{arcsec}/\text{century}$ for Mercury~\citep[see also][]{HD80606b_C11}.

Besides Mercury, the relativistic precession has been measured in the solar system for some asteroids like Icarus~\citep{Icarus_G53}. The tests done in the solar system agree with GR to within $10^{-4}$~\citep{Will}. Outside the solar system the effect is well known in binary pulsars like the Hulse-Taylor pulsar~\citep{HulseTaylor} where it has been measured with high precision~\citep[see][]{TW82}, and in other binary pulsars such as the double pulsar~\citep{DoublePulsar}. To a further distance, there are now attempts to measure the relativistic precession of the star S2 close to the galactic centre~\citep{S02_P17}. The measure of the relativistic periastron advance in stellar systems was proposed by~\citet{HD80606b_Gi85}, but it is quite complicated to discriminate it from the tidal precession (see \textit{e.g.}~\citet{HD80606b_Wo10}).

\citet{HD80606b_F16} reported that the relativistic precession was not detectable in the HD\,80606 system in spite of the large dataset secured over several years. They measured the variation of the longitude of the periastron using the whole available datasets: $\dot \omega = 9\,720 \pm 11\,160\,\text{arcsec}/\text{century}$, thus in a non-significant way. Indeed, the accuracy reached on $\omega$ mainly from radial velocities is $\pm\,540\,\text{arcsec}$~\citep{HD80606b_H11}, which is not accurate enough.

The transiting nature of HD\,80606b allows a better accuracy to be reached thanks to the precise timing that can be measured on the mid-point of each transit or eclipse. Typical accuracies of $\pm\,85\,\text{sec}$~\citep{HD80606b_H11} and $\pm\,260\,\text{sec}$~\citep{HD80606b_LDL09} were  obtained on the mid-point of transits and eclipses, respectively, using the \textit{Spitzer Space Telescope} and its IRAC instruments. Indeed, only a satellite on an Earth-trailing heliocentric orbit such as \textit{Spitzer} could allow the whole 12-hour-long transit of HD\,80606b to be continuously observed together with sufficiently long off-transit references immediately before and after the event. 

In this paper we propose to detect the GR precession of the periastron (and the relativistic orbital motion) for the exoplanet HD\,80606b by using a method based on the transit times, computing a small drift in the successive instants of transits due to relativity. We find that in ten years (after 33 periods), due to the relativistic effects, the instants of transits are delayed with respect to the Newtonian predicted ones by about three~minutes. Whereas the transit mid-time could be measured with such an accuracy, the absolute value of that shift could not be detected because the system's orbital parameters are not known today with a high enough accuracy.

However, we argue that the effect could actually be measurable from the observation of the change of the elapsed time $t_\text{tr$-$ec}=t_\text{tr}-t_\text{ec}$ between an eclipse and the next transit.\footnote{While in the final stage of the preparation of this paper, we found that this method was already suggested by~\citet{HD80606b_JB09,HD80606b_PK08} and later by~\citet{HD80606b_Io11}. Those works assume a central transit and use only the effect of the periastron shift. Furthermore they are generic while here we also investigate in detail the remarkable case of HD\,80606b.} The time between an eclipse and a transit was measured in January 2010 using \textit{Spitzer} to be
\begin{equation}
t_\text{tr$-$ec}(\text{Jan. 2010}) \simeq 5.9\,\text{days}\,,
\end{equation}
with a good accuracy of $\pm\,275\,\text{sec}$. Due to the GR precession of the periastron, and the modification of the period due to relativity, we find that this time difference will be reduced by $183\,\text{sec}$ in 2020 (after 33 orbits), and by $271\,\text{sec}$ in 2024 (49 orbits). Such shift would be difficult to detect in 2020 using IRAC on \textit{Spitzer}, but should be detectable a few years later using NIRCam or MIRI on \emph{James Webb Space Telescope} (JWST).

Most of the paper is devoted to theoretical aspects, and it ends with a conclusion and discussion in Sect.~\ref{sec:concl} where we give further comments on previous observations and on the Newtonian perturbing effects.
In Sect.~\ref{sec:geom} we introduce the geometrical conventions used to compute the times of transits and eclipses. Those are computed in two different ways: a post-Keplerian parametrisation of the orbit in Sect.~\ref{sec:PPK}, and a Hamiltonian method using Delaunay-Poincar\'e canonical variables in Sect.~\ref{sec:Ham}. Those methods give the exact solution of the first post-Newtonian equations of motion. Our final results for the transit times are given in Tables~\ref{table1},~\ref{table2}, and~\ref{table3} below. Furthermore, as a check of the latter methods, we also computed the transit times with a method relying on celestial-mechanics perturbation theory in Sect.~\ref{sec:lag}. As we used it, this procedure is valid on average over one orbit, and takes only into account the secular effects. We find that it is in good agreement with the post-Keplerian and the Hamiltonian methods. A quick but rough estimate of the GR shift of transit times is also given in Appendix~\ref{app:rough}.

\section{Geometry of the planetary transits} 
\label{sec:geom}

\begin{figure}[t]
    \centering
    \subfloat[][]{
        \includegraphics[scale=.4]{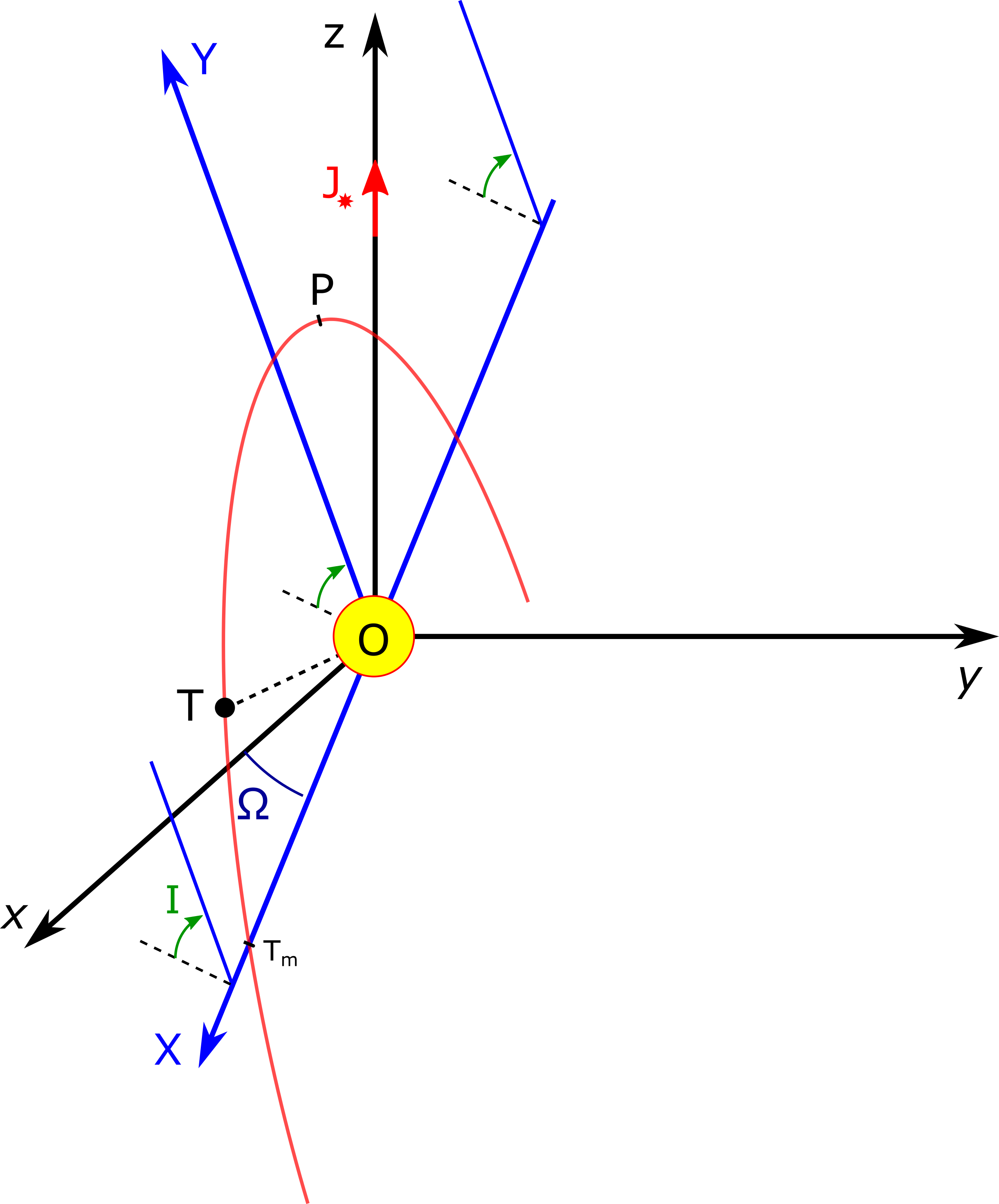}
        \label{fig_referentiel}}
    \hspace{.1cm}
    \subfloat[][]{
        \includegraphics[scale=.4]{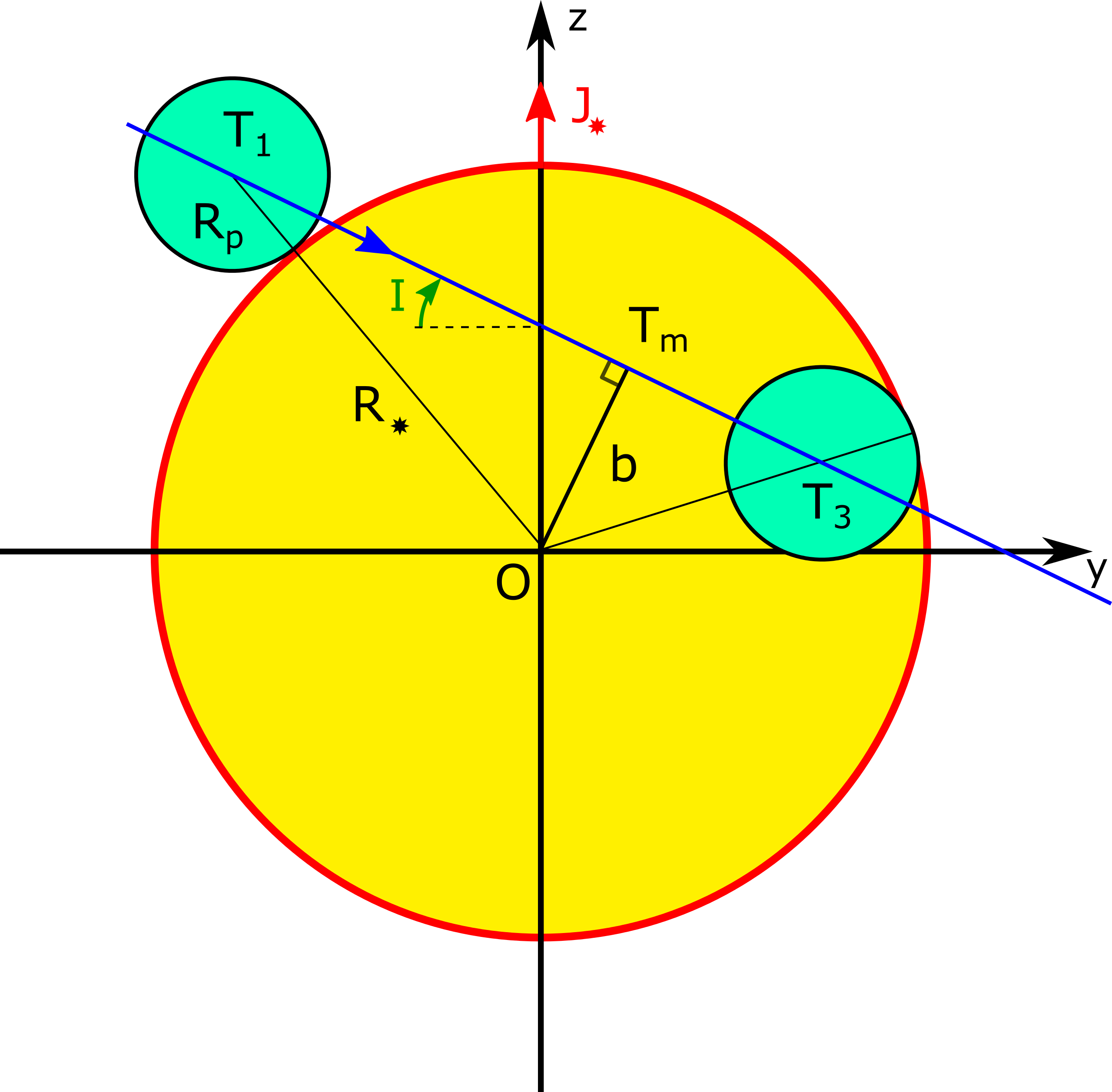}
        \label{fig_transit_face}}\\
    \subfloat[][]{
        \includegraphics[scale=.6]{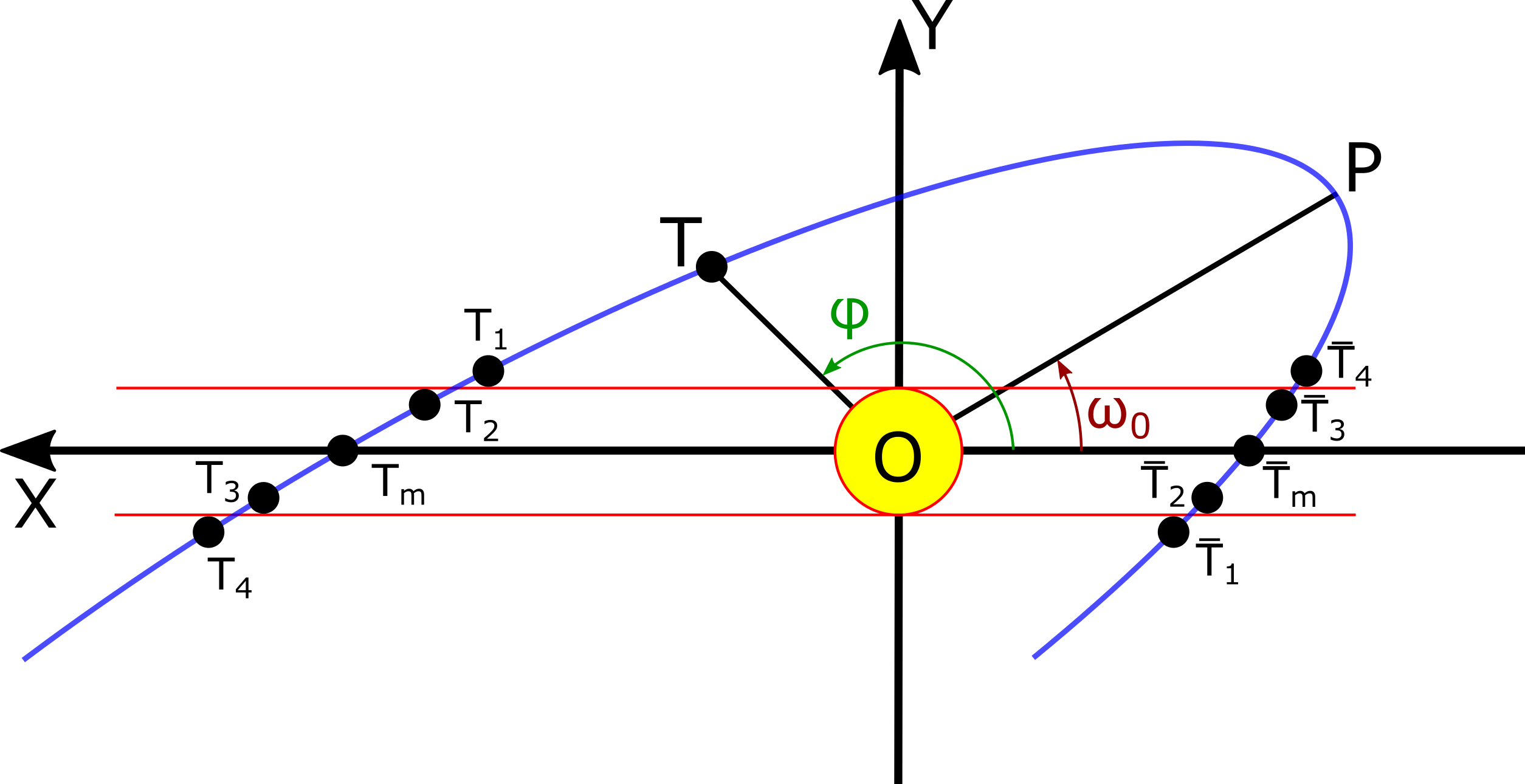}
        \label{fig_transit_cote}}
    \hspace{.1cm}
    \subfloat[][]{
        \includegraphics[scale=.4]{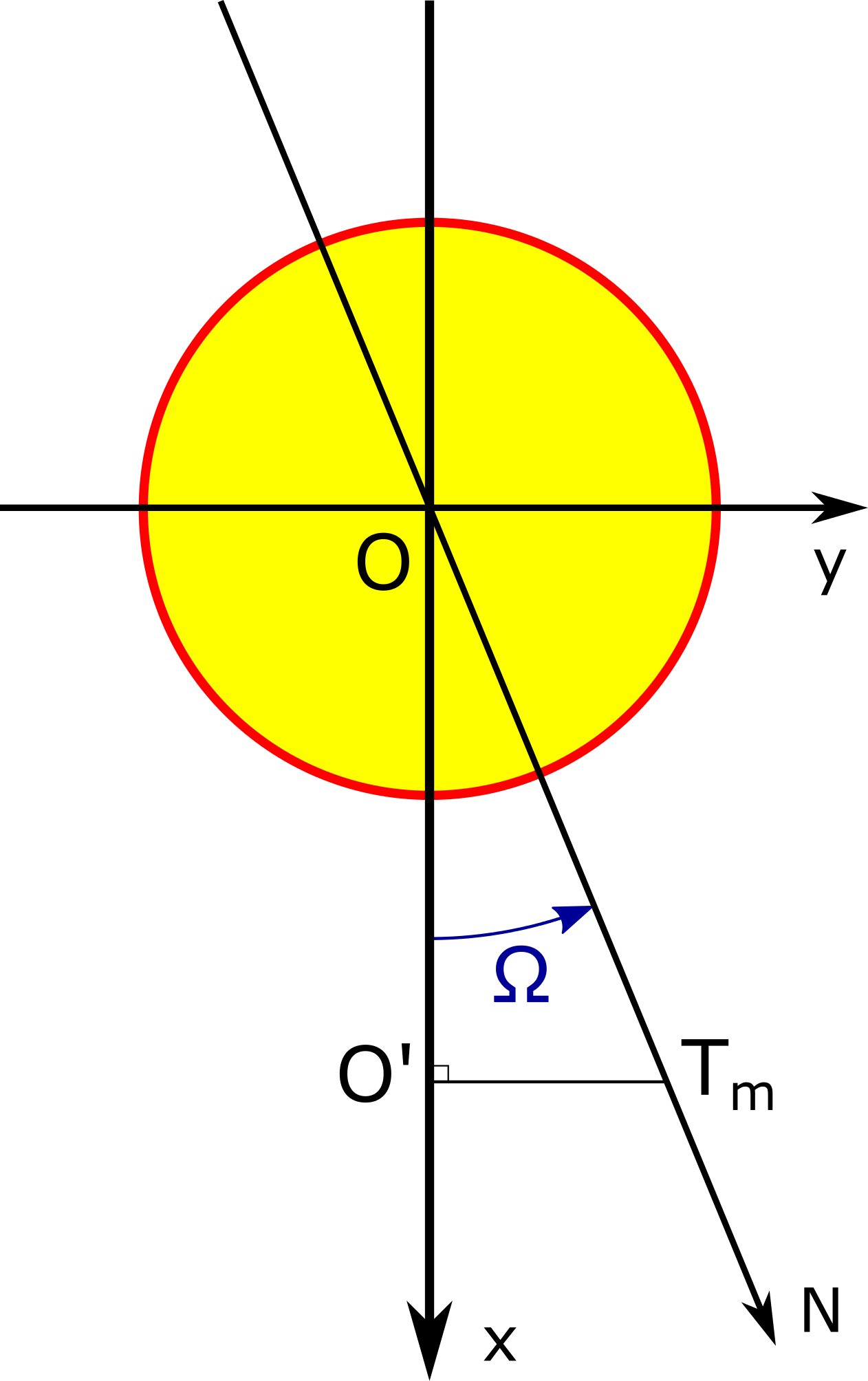}
        \label{fig_transit_haut}}    
    \caption{Geometry of the planetary transit:
    \protect\subref{fig_referentiel} Motion of the planet with respect to the observer, located at infinity in the direction of $x$ of the frame $\{x, y, z\}$. The centre of the star is denoted $O$, $P$ is the periastron at reference time $t_P,$ and $J_\star$ is the angular momentum of the star projected on the plane of the sky $\{y, z\}$.
    \protect\subref{fig_transit_face} The transit as seen by the observer and the definition of the points $\{T_i\}$; here are only shown $T_1$ and $T_3$.
    \protect\subref{fig_transit_cote} The trajectory of the planet and the points $\{T_i\}$ (transit) and $\{\bar{T}_i\}$ (eclipse) as seen in the orbital plane $\{X,Y\}$. Up to an angle $\Omega$, the observer is in the direction of $X$.
    \protect\subref{fig_transit_haut} View of the transit from above. The direction $(ON)$ is the line of ``nodes'', corresponding to $(OX)$ in \protect\subref{fig_referentiel}.}
    \label{fig_transit}
\end{figure}
We introduce a frame $\{x, y, z\}$ with origin at the centre of the star. The observer is in the direction $x$ (Earth's direction), while the projected angular momentum of the star is by convention along the direction $z$. Hence the plane $\{y, z\}$ is the plane of the sky. The orbital plane of the planet is defined by its inclination $I$ with respect to the plane $\{x, y\}$, and by its longitude $\Omega$ with respect to the plane $\{x, z\}$, following our conventions described in Fig.~\ref{fig_referentiel}. As depicted in Fig.~\ref{fig_transit_cote}, the motion of the planet in the orbital plane is parametrised by polar coordinates $(r,\varphi)$, where $\varphi$ denotes the sum of the true anomaly and of $\omega_0$, the latter denoting a constant angle, which can be viewed as defining the initial argument of the periastron at a reference time. Then the coordinates $(x, y, z)$ of the planet in this frame read (see Figs.~\ref{fig_referentiel} and~\ref{fig_transit_cote})
\begin{subequations}\label{xyz}
\begin{align}
x &= r \bigl[ - \cos\varphi \cos\Omega + \sin\varphi\cos I\sin\Omega\bigr]\,,\\
y &= r \bigl[ - \cos\varphi\sin\Omega - \sin\varphi\cos I\cos\Omega\bigr]\,,\\
z &= r \sin\varphi\sin I\,.
\end{align}
\end{subequations}
Since $x$ is the direction of the observer, the condition for the (partial or complete) transit of the planet is that $y^2+z^2\leqslant (R_\star+R_p)^2,$ where $R_\star$ and $R_p$ are the radius of the star and planet, hence
\begin{equation}\label{transit}
\sin^2 I\sin^2\varphi + \bigl(\cos I \cos\Omega\sin\varphi + \sin\Omega\cos \varphi\bigr)^2 \leqslant \left(\frac{R_\star+R_p}{r}\right)^2\,.
\end{equation}
Since $R_\star+R_p\ll r,$ it is easy to see that this condition can be satisfied when $\sin I\ll 1$ or $\sin\varphi\ll 1$, and $\vert\cos I \cos\Omega\sin\varphi + \sin\Omega\cos\varphi\vert\ll 1$. With the conventions of Fig.~\ref{fig_referentiel} we are interested in a transit for which $\varphi\simeq \pi$ and $\Omega \simeq 0$.\footnote{For HD\,80606b, \citet{HD80606b_H11} gave $\sin\Omega \simeq 0.019$.}
On the other hand, for an eclipse during which the planet is passing beyond the star as seen from the observer, we have $\varphi \simeq 0$ and $\Omega \simeq 0$. In an expansion to the first order in $\frac{R_\star+R_p}{r}\ll 1,$ we find that the coordinates of the planet during and around the transit are
\begin{subequations}\label{xyztransit}
\begin{align}
x &\simeq \frac{b}{\sin I} \Bigl[\cot\Omega +\cos I\,\sin\varphi\Bigr]\,,\\
y &\simeq \frac{b}{\sin I}\Bigl[ 1 - \frac{\sin\varphi\cos I}{\sin\Omega}\Bigr]\,,\\
z &\simeq\frac{b \,\sin\varphi}{\sin\Omega}\,,
\end{align}
\end{subequations}
where the impact parameter $b$ of the planet is the distance between the trajectory of the planet projected on the plane $\{y, z\}$ and the centre of the star.
Neglecting the bending of the trajectory during the transit and eclipses,\footnote{The correction induced by taking into account this effect is of the order of $\frac{\delta b}{b} \sim \frac{1}{2}\left(\frac{R_\star}{r}\right)^2$, so $\sim 10^{-4}$ for the transits and $\sim 10^{-2}$ for the eclipses, in the case of HD\,80606b.} the impact parameter is given by (see Fig.~\ref{fig_transit_face})
\begin{equation}\label{b}
b = \left\{\begin{array}{lll} r(\pi)\sin\Omega \sin I &&\text{(transits $T_i$)}\,,\\[0.4cm] r(0)\sin\Omega \sin I &&\text{(eclipses $\bar{T}_i$)}\,,
\end{array}\right.
\end{equation}
where $r(\pi)$ and $r(0)$ are the values of the radial coordinate of the planet at $\varphi=\pi$ and $\varphi=0$, when the transit and eclipse occur. Since the eclipse will be seen from behind the star and so much farther away, we shall have to add to the instants of eclipse a Roemer time delay accounting for the propagation of light at finite velocity $c$, and which is a 0.5PN effect $\propto 1/c$. Thus $\bar{t}_i \longrightarrow \bar{t}_i + \Delta t_\text{R}$ where  
\begin{equation}\label{Roemer}
\Delta t_\text{R} = \frac{r(0) + r(\pi)}{c}\, \simeq 2.7 \;\text{min $\quad$(for
HD\,80606b)}\,.
\end{equation}

We define the positions of the planet $\{T_1, T_2, T_3, T_4\}$ to be respectively the entry of the transit, the start of the full transit, and the exits of the full and partial transits. Similarly the points $\{\bar{T}_1, \bar{T}_2, \bar{T}_3,\bar{T}_4\}$ corresponding to the eclipse are shown in Fig.~\ref{fig_transit_cote}. The transit and eclipse conditions for the polar coordinates $(r,\varphi)$ at the passage of these points read
\begin{equation}\label{geomcondi}
r \sin\varphi = Y(b)\,.
\end{equation}
Given a model $r(\varphi)$ for the planetary orbit, this determines the true anomaly $\varphi$ for each of these points, where the vertical coordinate $Y(b)$ is given in terms of the impact parameter $b$ by (see Figs.~\ref{fig_transit_face} and~\ref{fig_transit_cote})
\begin{equation}\label{Yb}
Y(b) = \left\{\begin{array}{lll}
\sqrt{(R_\star+R_p)^2-b^2} + b\,\cot I &&\text{($T_1$ and $\bar{T}_4$)}\,,\\[0.4cm]
\sqrt{(R_\star-R_p)^2-b^2} + b\,\cot I &&\text{($T_2$ and $\bar{T}_3$)}\,,\\[0.4cm]
-\sqrt{(R_\star-R_p)^2-b^2} + b\,\cot I &&\text{($T_3$ and $\bar{T}_2$)}\,,\\[0.4cm]
-\sqrt{(R_\star+R_p)^2-b^2} + b\,\cot I &&\text{($T_4$ and $\bar{T}_1$)}\,.
\end{array}\right.
\end{equation}
The cases for which $(R_\star \pm R_p)^2=b^2$ correspond to the planet just grazing tangentially the star, either from the exterior or from the interior of the stellar disc. In addition to~Eqs. \eqref{geomcondi}--\eqref{Yb} we compute also the points corresponding to the minimal approach to the center of the star, $T_m$ and $\bar{T}_m$, whose transit conditions are obviously (see Fig.~\ref{fig_transit_face})
\begin{equation}\label{Tm}
Y(b) = b\,\cot I \qquad\text{($T_m$ and $\bar{T}_m$)}\,.
\end{equation}
Having determined the polar coordinates of the planet we shall deduce the instants $t_i$ and $\bar{t}_i$ of passage at each of these points by using the law of motion of the planet on the trajectory.

\section{Post-Newtonian motion of the planet}
\label{sec:PPK}

In this section we compute the times of transits and eclipses, that is, the instants of the successive passages at the positions $T_i$ and $\bar{T}_i$ defined by the geometric conditions in Eqs.~\eqref{geomcondi}--\eqref{Tm}, in a relativistic model, including the first post-Newtonian (1PN) corrections beyond a Keplerian orbit. More precisely, we compute the difference between the transit times predicted by the relativistic 1PN model and those predicted by the Newtonian model. This will permit us to assess whether the latter relativistic corrections could be detectable, and we propose an observable quantity, directly measurable in future observations of the planet HD\,80606b.

\subsection{Keplerian parametrisation of the orbit}

In the Keplerian model the orbit of the planet and its motion on the orbit are given by the usual six orbital elements $\{a, e, I, \ell, \omega_0, \Omega\}$, where $I$ and $\Omega$ specify the orbital plane (see Fig.~\ref{fig_referentiel}), $a$ is the semi-major axis, $e$ is the eccentricity, $\omega_0$ denotes the constant angular position of the periastron, and the mean anomaly $\ell$ is defined by $\ell=n_0(t-t_{0,P}),$ where $t_{0,P}$ is the instant of passage at the periastron and $n_0=2\pi/P_0$ is the mean motion, with $P_0$ the period. The Keplerian relative orbit in polar coordinates $(r,\varphi)$, is conveniently parametrised by the eccentric anomaly $\psi$ as
\begin{subequations}\label{Kepler}
\begin{align}
r &= a \bigl( 1 - e \cos\psi \bigr) \,,\\
\ell &= \psi - e \sin\psi\,,\label{eqkepler}\\
\varphi &= \omega_0 + 2 \arctan \biggl(\sqrt{\frac{1+e}{1-e}} \tan \frac{\psi}{2}
\biggr)\,.
\end{align}
\end{subequations}
The relative orbit in polar coordinates reads
\begin{equation}\label{n}
r = \frac{a(1-e^2)}{1+ e \cos(\varphi-\omega_0)}\,,
\end{equation}
from which we deduce the impact parameter in the Newtonian model, say $b_0$, to be
\begin{equation}\label{b0}
b_0 = \left\{\begin{array}{lll} \displaystyle a(1-e^2)\frac{\sin\Omega \sin I}{1-e \cos\omega_0} &&\text{(transits $T_i$)}\,,\\[0.4cm] \displaystyle
a(1-e^2)\frac{\sin\Omega \sin I}{1+e \cos\omega_0} &&\text{(eclipses $\bar{T}_i$)}\,.
\end{array}\right.
\end{equation}
In the Newtonian model we have Kepler's third law $n_0=(G M/a^3)^{1/2}$, where $M=M_\star+M_p$ is the total mass of the star plus planet system. Furthermore the energy and the angular momentum of the orbit are given by the usual Keplerian formulas
\begin{subequations}\label{EJ}
\begin{align}
E &= - \frac{G M\mu}{2 a}\,,\\J &= \mu \sqrt{G M a (1-e^2)}\,,
\end{align}
\end{subequations}
where $\mu=M_\star M_p/M$ is the reduced mass. 

The conditions for the transits and eclipses are given by Eqs.~\eqref{geomcondi} with~\eqref{Yb}, and determine the values of the eccentric anomaly at these points.
In the Newtonian analysis the eccentric anomaly $\psi_0$ is obtained by solving 
\begin{equation}\label{cond0}
a \left[ \sqrt{1-e^2} \sin\psi_0 \cos\omega_0 + (\cos\psi_0-e)\sin\omega_0\right] =
Y_0\,,
\end{equation}
where $Y_0=Y(b_0)$ denotes the vertical coordinate~(Eq. \eqref{Yb}) computed with the Newtonian impact parameter given by~Eq. \eqref{b0}.

\subsection{Relativistic corrections to the Keplerian motion}

Next we consider the relativistic 1PN model~\citep[see \emph{e.g.}][]{WagW76}. For this we shall use an explicit solution of the equations of motion that is particularly elegant, called the quasi-Keplerian representation of the 1PN motion~\citep{DD85}. In this representation the orbit and motion are parametrised by the eccentric anomaly $\psi$ in a Keplerian-looking form as
\begin{subequations}\label{QKepler}
\begin{align}
r &= a_r \bigl( 1 - e_r \cos\psi \bigr) \,,\\
\ell &= \psi - e_t \sin\psi\,,\\\varphi &= \omega_0 + 2 K \arctan
\biggl(\sqrt{\frac{1+e_\varphi}{1-e_\varphi}} \tan \frac{\psi}{2} \biggr)\,,
\end{align}
\end{subequations}
where $\ell=n(t-t_P)$ and $n=2\pi/P$, with now $P$ being the period of the relativistic model. The motion is parametrised by six constants: the mean motion $n$, a particular definition of semi-major axis $a_r$ differing from $a$ in the Newtonian model by a small 1PN term, the precession of the orbit $K$ including the relativistic precession, and three types of eccentricities $e_r$, $e_t,$ and $e_\varphi$ differing from $e$ and from each other by small 1PN terms. We shall pose
\begin{subequations}\label{defparameters}
\begin{align}
& n= n_0\bigl(1+\zeta\bigr)\,, & K = 1+k \,, & \qquad\qquad a_r =
a\bigl(1+\xi\bigr)\,,\\
& e_r = e + \varepsilon_r \,, & e_t = e + \varepsilon_t \,, & \qquad\qquad e_\varphi
= e + \varepsilon_\varphi \,.
\end{align}
\end{subequations}

The constant $k$ denotes the relativistic advance of the periastron \textit{per} orbital revolution, which is $k=\frac{\Delta}{2\pi}$ where $\Delta$ is the angle of return to the periastron, and whose GR value has been given in~Eq. \eqref{DeltaGR}. The effect of precession can be seen clearly with the expression of the relative orbit in polar coordinates, which takes the more involved form~\citep{DD85}
\begin{equation}\label{orbitprecession}
r = \frac{a_r(1-e_r^2)}{1+ e_r \cos\bigl[\frac{\varphi-\omega_0}{K}-\frac{1}{6}k \,e_r \,\nu \sin\bigl(\frac{\varphi-\omega_0}{K}\bigr)\bigr]}\,.
\end{equation}
Besides the usual contribution of the precession $\propto\frac{\varphi-\omega_0}{K}$, we note the presence of a term vanishing in the small mass ratio limit, when $\nu=\mu/M\to 0$. 

Crucial to the post-Keplerian formalism are the explicit expressions of all these constants in terms of the conserved energy $E$ and angular momentum $J$ of the orbit. Since we want to compare the relativistic model~of Eq. \eqref{QKepler} to the Newtonian one~of Eq. \eqref{Kepler} we will compare, for convenience, orbits that have the same energy $E$ and angular momentum $J$. Thus we need to express the relations in terms of the ``Newtonian'' semi-major axis $a$ and eccentricity $e$ defined in terms of $E$ and $J$ by the Newtonian formulas~in Eq. \eqref{EJ}. We have~\citep{DD85}~\citep[see also][]{BlanchetLR}
\begin{subequations}\label{parameters}
\begin{align}
\zeta &= \frac{G M}{8a c^2}\bigl(-15+\nu\bigr)\,,\label{zeta}\\
k &= \frac{3G M}{a c^2(1-e^2)}\,,\label{k}\\
\xi &= \frac{G M}{4a c^2}\bigl(-7+\nu\bigr)\,,\\
\varepsilon_r &= \frac{G M}{8a
c^2}\biggl[\frac{9+\nu}{e}+\bigl(15-5\nu\bigr)e\biggr] \,,\\
\varepsilon_t &= \frac{G M}{8a
c^2}\biggl[\frac{9+\nu}{e}+\bigl(-17+7\nu\bigr)e\biggr]\,,\\
\varepsilon_\varphi &= \frac{G M}{8a
c^2}\biggl[\frac{9+\nu}{e}+\bigl(15-\nu\bigr)e\biggr] \,.
\end{align}
\end{subequations}
Among the effects described by these constants, the main ones are those associated with the mean motion $n$ and the periastron precession $K$. The expressions for $n$ and $K$ in terms of $E$ and $J$ are invariant, meaning they do not depend on the coordinate system. Since
\begin{equation}\label{ksurzeta}
\frac{k}{\zeta} = -\frac{24}{(15-\nu)(1-e^2)} \simeq -12 \quad\text{(for
HD\,80606b)}\,,
\end{equation}
we expect the effect of the precession of the orbit to be dominant over the shift in
the orbital period. The other relations, linking the semi-major axis and
eccentricities to $E$ and $J$, depend on the coordinate system; they are given here
in harmonic coordinates. In this paper we compute an observable quantity
(defined by Eq.~\eqref{DeltatN} below), which thus does not depend on the choice of
coordinate system.

We now investigate the difference between the analysis using the relativistic model and the Newtonian one concerning the planetary transits. The conditions for the transits and eclipses are still given by Eqs.~\eqref{geomcondi} with~\eqref{Yb}, with the impact parameter given by~Eq. \eqref{b}. However, the impact parameter will differ from the Newtonian value by an amount $\delta b$. We readily find this amount thanks to the explicit formula~Eq.~\eqref{orbitprecession}, using a first order expansion in the small parameters given by Eq.~\eqref{parameters},
\begin{equation}\label{deltab}
\delta b = 
\left\{\begin{array}{llll} \displaystyle b_0\Bigl[\xi-\frac{2 e\,
\varepsilon_r}{1-e^2}+\frac{\varepsilon_r \cos\omega_0}{1-e \cos\omega_0}
-\frac{ke\sin\omega_0}{1-e \cos\omega_0}\left(\varphi_N +
\frac{e\nu\sin\omega_0}{6}\right)\Bigr] &&&\text{(transits $T_i$)}\,,\\[0.5cm]
%%%%%%%%%%%%%%%%%%%%%%%%%%%%%%%%%%%%%%%%%%%%%%%%%%%%%
\displaystyle b_0\Bigl[\xi-\frac{2 e\, \varepsilon_r}{1-e^2}-\frac{\varepsilon_r
\cos\omega_0}{1+e \cos\omega_0} + \frac{ke\sin\omega_0}{1+e
\cos\omega_0}\left(\bar\varphi_N - \frac{e\nu\sin\omega_0}{6}\right)\Bigr]
&&&\text{(eclipses $\bar{T}_i$)}\,,
\end{array}\right.
\end{equation}
where $\varphi_N = (2N+1)\pi-\omega_0$ and $\bar\varphi_N = 2N\pi-\omega_0$. Here $N$ is the number of orbits since the reference time, with $N=0$ corresponding to the transit observed in January 2010. Then the coordinate $Y(b)$ is modified by $\delta Y=\frac{\dd Y}{\dd b}\vert_{b_0}\,\delta b$ and we find (see Eq.~\eqref{Yb}),\footnote{When $(R_\star \pm R_p)^2=b^2$ the 1PN perturbation of the Newtonian transit condition breaks down since the denominator of~\eqref{deltaY} is zero. This corresponds to the case where the planet is just grazing the disc of the star either from inside or outside. For HD\,8606b we have $b_0\simeq 0.73\,(R_\star+R_p)$and $b_0\simeq 0.90\,(R_\star-R_p)$ so we are far from this case.}
\begin{equation}\label{deltaY}
\delta Y = \frac{Y_0 \sin(2 I) - 2 b_0}{2\,Y_0 \sin^2 I - b_0 \sin(2I)} \,\delta b\,.
\end{equation}
% % 
Naturally for $T_m$ and $\bar{T}_m$, Eq.~\eqref{Tm}, we just have $\delta Y = \delta b \cot I$.

We are now in a position to solve the transit conditions and find the modification of the eccentric anomaly $\delta\psi$ with respect to the Newtonian model, that is, the solution of Eq.~\eqref{cond0}. Among all the relativistic effects contributing to $\delta\psi$, only the one associated with the relativistic precession $k$ is ``secular'', and therefore grows when $\psi_0\longrightarrow\psi_0+2\pi N$ proportionally to the number of orbits $N$. The other effects, associated with the relativistic parameters $\xi$, $\varepsilon_r,$ and $\varepsilon_\varphi$, are periodic and therefore average to zero in the long term. Accordingly we split
\begin{equation}
\delta\psi = \delta\psi_\text{secular} + \delta\psi_\text{periodic}\,.
\end{equation}
Similarly, we distinguish $\delta Y = \delta Y_\text{secular} + \delta Y_\text{periodic}$, where the secular part is given by the $N$-dependent terms in~Eq. \eqref{deltab}. Perturbing Eq.~\eqref{geomcondi} around the solution $\psi_0$ of the Newtonian condition~of Eq. \eqref{cond0}, we explicitly find
\begin{subequations}\label{deltaPsiexpl}
\begin{align}
\delta\psi_\text{secular} &= 2 k \frac{(\cos\psi_0-e)\cos\omega_0 - \sqrt{1-e^2}
\sin\psi_0 \sin\omega_0}{\sin\psi_0 \sin\omega_0 - \sqrt{1-e^2} \cos\psi_0
\cos\omega_0}\arctan \biggl(\sqrt{\frac{1+e}{1-e}} \tan \frac{\psi_0}{2}
\biggr)\nonumber\\
&  - \frac{a^{-1}\delta Y_\text{secular}}{\sin\psi_0 \sin\omega_0 - \sqrt{1-e^2}
\cos\psi_0 \cos\omega_0}\,,\\
\delta\psi_\text{periodic} &= \frac{(\cos\psi_0-e)\sin\omega_0 + \sqrt{1-e^2}
\sin\psi_0 \cos\omega_0}{\sin\psi_0 \sin\omega_0 - \sqrt{1-e^2} \cos\psi_0
\cos\omega_0}\left( \xi + (\varepsilon_\varphi-\varepsilon_r)\frac{\cos\psi_0}{1-e
\cos\psi_0}\right) \nonumber\\
& - \frac{e (1-e^2)^{-1/2} \varepsilon_\varphi \sin\psi_0 \cos\omega_0 +
\varepsilon_\varphi \sin\omega_0 + a^{-1}\delta Y_\text{periodic}}{\sin\psi_0
\sin\omega_0 - \sqrt{1-e^2} \cos\psi_0 \cos\omega_0}\,.
\end{align}
\end{subequations}
Finally, having computed $\delta\psi$ with the transit condition we readily obtain the time difference between the instants of transits (and eclipses) in the relativistic model and in the Newtonian model by using the equation for the mean anomaly $\ell = n(t-t_P) = \psi-e_t \sin\psi$. We notice that there is another secular term associated with the correction in the period, $n= n_0(1+\zeta)$, where $\zeta$ is the second invariant of the motion besides the relativistic precession $k$. We thus write similarly $\delta t = \delta t_\text{secular} + \delta t_\text{periodic}$ and find
\begin{subequations}\label{deltat}
\begin{align}
\delta t_\text{secular} &= \frac{1}{n_0}\bigl[(1- e
\cos\psi_0)\delta\psi_\text{secular} - \zeta (\psi_0 - e \sin\psi_0)\bigr]\,,\\
\delta t_\text{periodic} &= \frac{1}{n_0}\bigl[(1- e
\cos\psi_0)\delta\psi_\text{periodic} - \varepsilon_t \sin\psi_0 \bigr]\,.
\end{align}
\end{subequations}

When computing $\delta \bar{t}$, the 1.5PN shift in the Roemer effect also has to be taken in account, although we find that this effect is very small. With the help of Eq.~\eqref{Roemer}, it comes after $N$ orbits
\begin{align}\label{Roemer_pert}
\delta t_\text{R} = 
\frac{2a(1-e^2)\xi}{c^3(1-e^2\cos^2\omega_0)}
-\frac{4a\,e\,\varepsilon_r}{c^3(1-e^2\cos^2\omega_0)^2} -\frac{a
k\,(1-e^2)e\sin\omega_0}{c^3\left(1-e\cos\omega_0\right)^2}\left[\pi +
4\frac{\bar\varphi_N e\cos\omega_0}{(1+e\cos\omega_0)^2}
+\frac{\nu(1+e^2\cos^2\varphi)e\sin\omega_0}{3(1+e\cos\omega_0)^2}\right]\,.
\end{align}

Finally let us comment on the approximation we have made here, namely the expansion to first order in the relativistic parameters $k$, $\zeta$, $\xi$, and so on. This approximation is valid as long as the number of orbits $N$ over which we integrate is much smaller than the inverse of these relativistic effects, for example as long as $N\ll 1/k$. In the case of HD\,80606b we have $k \simeq 5\cdot 10^{-7}$ and we are considering $N = 33$ orbits between the transit of January 2010 and that of 2020 (see the Introduction), so our approximation is justified.

We present our results\footnote{We took our values in Table 1 from \citet{HD80606b_H11}. We note that our angular conventions in
Fig.~\ref{fig_referentiel} are different: $I = \lambda$ (hence $I = 42^{\,\circ}$ adopted here) and $\sin\Omega \simeq \frac{\cos i}{\sin\lambda}$.} in Tables~\ref{table1},~\ref{table2}, and~\ref{table3}. From the above analysis we compute for each orbit $N$ the quantities (with $i\in\{1,2,\text{m},3,4\}$)
\begin{subequations}\label{deltatiN}
\begin{align}
\delta t_i(N) &= t_i(N) - t_{0,i}(N) \,,\\
\delta \bar{t}_i(N) &= \bar{t}_i(N) - \bar{t}_{0,i}(N) \,,
\end{align}
\end{subequations}
where $t_{0,i}(N)$ and $\bar{t}_{0,i}(N)$ denote the Newtonian values. Obviously we have $t_{0,i}(N) = t_{0,i}(0)+N P_0$ and $\bar{t}_{0,i}(N) = \bar{t}_{0,i}(0)+N P_0$ since the Newtonian orbit is fixed, and the motion periodic with period $P_0=2\pi/n_0$. Together with the shift of instants of transit, we compute also the relativistic effect on the total duration and the entrance of the $N$-th transit, as defined by
\begin{subequations}\label{deltat14N}
\begin{align}
\delta t_{14}(N) &= \delta t_4(N) - \delta t_1(N) \,,\\
\delta t_{12}(N) &= \delta t_2(N) - \delta t_1(N) \,,
\end{align}
\end{subequations}
together with similar quantities $\delta \bar{t}_{14}(N)$ and $\delta
\bar{t}_{12}(N)$ for the $N$-th eclipse. The results are given in
Tables~\ref{table1} and~\ref{table2}.

\begin{table*}[h]
        \begin{center}
                \begin{tabular}{| l || c | c | c |c | c || c | c |}
                        \hline\hline
                        number $N$ of transit & $\delta t_1(N)$ & $\delta t_2(N)$ & $\delta
t_m(N)$ & $\delta t_3(N)$ & $\delta t_4(N)$ & $\delta t_{14}(N)$& $\delta
t_{12}(N)$\\
                        \hline
                        0 (January 2010) & $-2.65$ & $-2.78$ & $-2.73$ & $-2.49$ & $-2.62$ & $0.04$ &
$-0.13$ \\
                        1 (May 2010) & $-7.70$ & $-8.08$ & $-7.94$ & $-7.24$ & $-7.61$ & $0.09$ & $-0.38$ \\
                        2 (August 2010) & $-12.76$ & $-13.38$ & $-13.16$ & $-11.99$ & $-12.60$ & $0.15$ &
$-0.62$ \\
                        \hspace{1cm} $\vdots$ & $\vdots$ & $\vdots$ & $\vdots$ & $\vdots$ & $\vdots$ &
$\vdots$ & $\vdots$\\
                        32 (October 2019) & $-164.3$ & $-172.4$ & $-169.5$ & $-154.4$ & $-162.4$ & $1.93$
& $-8.15$ \\
                        33 (February 2020) & $-169.3$ & $-177.7$ & $-174.7$ & $-159.1$ & $-167.4$ &
$1.99$ & $-8.40$ \\
                        \hspace{1cm} $\vdots$ & $\vdots$ & $\vdots$ & $\vdots$ & $\vdots$ & $\vdots$ &
$\vdots$ & $\vdots$\\
                        48 (September 2024) & $-245.1$ & $-262.6$ & $-252.9$ & $-235.0$ & $-247.2$ &
$2.88$ & $-12.4$ \\
                        49 (December 2024) &  $-250.2$ & $-262.6$ & $-258.1$ & $-235.0$ & $-247.2$ &
$2.94$ & $-12.4$ \\
                        \hline\hline
                \end{tabular}
                \caption{Differences (in seconds) between the instants of transit computed with
the relativistic (1PN) model and with the Newtonian model, taking as reference
point ($\delta t \equiv 0$) the passage at periastron just before the transit of
Jan. 2010.} \label{table1}
        \end{center}
\end{table*}
\begin{table*}[h]
        \begin{center}
                \begin{tabular}{| l || c | c | c |c | c || c | c |}
                        \hline\hline
                        number $N$ of eclipse & $\delta \bar{t}_1(N)$ & $\delta \bar{t}_2(N)$ & $\delta
\bar{t}_m(N)$ & $\delta \bar{t}_3(N)$ & $\delta \bar{t}_4(N)$ & $\delta
\bar{t}_{14}(N)$& $\delta \bar{t}_{12}(N)$\\
                        \hline
                        0 (January 2010) & $7.3\cdot10^{-3}$ & $7.3\cdot10^{-3}$ & $8.6\cdot10^{-3}$ &
$8.2\cdot10^{-3}$ & $8.3\cdot10^{-3}$ & $1.0\cdot10^{-3}$ & $6.0\cdot10^{-5}$\\
                        1 (April 2010) & $0.33$ & $0.33$ & $0.34$ & $0.34$ & $0.34$ & $6.8\cdot10^{-3}$ &
$6.0\cdot10^{-4}$\\
                        2 (August 2010) & $0.65$ & $0.65$ & $0.66$ & $0.66$ & $0.67$ & $1.3\cdot10^{-2}$
& $1.1\cdot10^{-3}$\\
                        \hspace{1cm} $\vdots$ & $\vdots$ & $\vdots$ & $\vdots$ & $\vdots$ & $\vdots$ &
$\vdots$ & $\vdots$ \\
                        32 (October 2019) &  $10.35$ & $10.36$ & $10.46$ & $10.51$ & $10.53$ & $0.19$ &
$1.7\cdot10^{-2}$ \\
                        33 (February 2020) &  $10.67$ & $10.69$ & $10.78$ & $10.84$ & $10.86$ & $0.19$ &
$1.8\cdot10^{-2}$ \\
                        \hspace{1cm} $\vdots$ & $\vdots$ & $\vdots$ & $\vdots$ & $\vdots$ & $\vdots$ &
$\vdots$ & $\vdots$ \\
                        48 (August 2024) &  $15.51$ & $15.54$ & $15.68$ & $15.77$ & $15.79$ & $0.28$ &
$2.6\cdot10^{-2}$ \\
                        49 (December 2024) &  $15.84$ & $15.86$ & $16.01$ & $16.12$ & $16.10$ & $0.29$ &
$2.6\cdot10^{-2}$ \\
                        \hline\hline
                \end{tabular}
                \caption{Differences (in seconds) between the instants of eclipse computed with
the relativistic (1PN) model and with the Newtonian model, taking as reference
point ($\delta \bar{t} \equiv 0$) the passage at periastron just after the eclipse
of Jan. 2010. We include the variation of the Roemer delay given
by~\eqref{Roemer_pert}.} \label{table2}
        \end{center}
\end{table*}
As we see in Table~\ref{table1}, the relativistic model predicts that after ten years the planet will begin the transit about three minutes earlier than the Newtonian prediction, and about 4.5 minutes earlier after 15 years. Given that the precision on the measured instants of transits can be as good as 85~seconds~\citep{HD80606b_H11}, this is already an indication that the relativistic effects might be measurable if the orbital parameters were particularly well known. However, the ten-years-long relativistic effect on the transit duration $\delta t_{14}$ and transit entrance $\delta t_{12}$ are on the order of seconds, and seem \textit{a priori} hardly detectable. Concerning the eclipse in Table~\ref{table2}, the effect is much smaller, of the order of ten seconds after ten years with respect to the Newtonian model. As the effect on the transit and the eclipse timings is different, this provides a way to directly detect it.

\begin{table*}[h]
\begin{center}
        \begin{tabular}{| c || c |}
                \hline\hline
                $N$ & ~$t_\text{tr$-$ec}(N) \!-\! t_\text{tr$-$ec}(0)$~\\
                \hline
                0  (January 2010) &  $0$\\
                1  & $-5.54$\\
                2  & $-11.08$\\
                \hspace{0.3cm}$\vdots$\hspace{0.3cm} &  $\vdots$\\
                32  (October 2019) &  $-177.3$\\
                33  (February 2020) &  $-182.8$\\
                \hline\hline
        \end{tabular}
        \end{center}
        \vspace{-0.5cm}
\begin{center}
%\hspace{-1.0cm}        
\begin{tabular}{|c||*{10}{c|} }
                \hline\hline
                $N$ & 30 &31 & 32 & 33 & 34 & 35 & 36 & 37 & 38 & 39 \\
                \hline                
                $t_\text{tr$-$ec}(N)-t_\text{tr$-$ec}(0)$ & $-166.2$ & $-171.7$ & $-177.3$ &
$-182.8$ & $-188.3$ & $-193.9$ & $-199.4$ & $-204.9$ & $-210.5$ & $-216.0$ \\
                \hline\hline
                $N$ & 40 & 41 & 42 & 43 & 44 & 45 & 46 & 47 & 48 & 49 \\
                \hline                
                $t_\text{tr$-$ec}(N)-t_\text{tr$-$ec}(0)$ & $-221.6$ & $-227.1$ & $-232.6$ &
$-238.2$ & $-243.7$ & $-249.3$ & $-254.8$ & $-260.3$ & $-265.9$ & $-271.4$ \\
                \hline\hline
        \end{tabular}
                \caption{Differences (in seconds) between the time interval $t_\text{tr$-$ec}(N)$ between the passage at the minimum approach point during the $N$-th eclipse and the passage at the minimum approach point during the $N$-th transit, and the time interval $t_\text{tr$-$ec}(0)$ at the reference period, when the eclipse and transit where measured in January 2010 with the \textit{Spitzer} satellite~\citep{HD80606b_H11}.
The values for the next five years are also given. The 30th transit occurred in March 2019 and the 49th will occur in December 2024.} \label{table3}
\end{center}
\end{table*}
Next we define $t_\text{tr$-$ec}(N)$ to be the time interval between the passage at the minimum approach point during the $N$-th eclipse and the passage at the minimum approach point during the $N$-th transit:
\begin{equation}\label{DeltatNdef}
t_\text{tr$-$ec}(N) = t_m(N) - \bar{t}_m(N) \,.
\end{equation}
By a straightforward calculation, using the fact that the Newtonian orbit is fixed, we obtain
\begin{equation}\label{DeltatN}
t_\text{tr$-$ec}(N) -t_\text{tr$-$ec}(0) = \delta t_m(N) - \delta
\bar{t}_m(N) - \delta t_m(0) + \delta \bar{t}_m(0) \,.
\end{equation}
The important point about this result is that the right-hand side of the equation represents the relativistic prediction, which is tabulated in Tables~\ref{table1} and~\ref{table2}, while the left-hand side is directly measurable: namely $t_\text{tr$-$ec}(0)$ has been measured with good precision between the eclipse and the transit of 2010, while future observation campaigns could allow $t_\text{tr$-$ec}(N)$ to be measured ten to fifteen years later. From Table~\ref{table3} we predict that the observations in 2020 will measure that the time difference between the transit and eclipse (with $N=33$) is shorter by 182.8 seconds (approximately three minutes), because of the relativistic effect, as compared to the one which was measured ten years ago in 2010. The values of this observable quantity given by~Eq. \eqref{DeltatN} are also presented for the transits of the next seven years in Table~\ref{table3}; in December 2024, the difference between the eclipse and transit (with $N=49$) is shorter by 271.4 seconds (4.5 minutes) as compared to the one measured in 2010.

The 2010 measurement yielded~\citep{HD80606b_LDL09,HD80606b_H11}
\begin{equation}\label{Deltat0}
t_\text{tr$-$ec}(0) = (5.8491 \pm 0.003) \,\text{days}\,,
\end{equation}
with good published uncertainty in the measurement, of the order of $4.5$ minutes, comparable to the relativistic effect we want to detect. Therefore we conclude that for the next observations of the transit/eclipse in the coming years, we should already be on the verge of detecting the relativistic effect. 

\section{Hamiltonian integration of the motion}
\label{sec:Ham}

At the 1PN order, the Hamiltonian of the relative motion of two point masses reads~\citep{BlanchetLR}
\begin{equation}\label{eqHam}
\frac{H}{\mu} = \frac{1}{2}\,P^2- \frac{GM}{R} +
\frac{1}{c^2}\left[\frac{3\nu-1}{8}\,P^4 -\frac{GM}{2R}\left(\nu
P_R^2+(3+\nu)P^2\right)+ \frac{G^2M^2}{2R^2}\right]\,,
\end{equation}
where $\bm{P} = \frac{\bm{P}_1+\bm{P}_2}{\mu}$ is the reduced linear momentum, which is conjugate to $\bm{X}$, and we pose $P_R = \bm{P}\cdot\bm{X}/R$ with $R =\vert\bm{X}\vert$, and $P^2=\bm{P}^2$. We shall use the Delaunay-Poincar\'{e} canonical variables. We start with the usual orbital elements $\{a,e,\ell,\omega\}$, where $a$ and $e$ are the semi-major axis and eccentricity, $\omega$ is the argument of the periastron, and $\ell=n(t-t_P)$ denotes the mean anomaly, where $n=\sqrt{G M/a^3}$ by definition.\footnote{We would like to emphasise that $n$ defined here and in Sect.~\ref{sec:lag} (perturbation theory) does not have the same meaning as in the quasi-Keplerian representation of the 1PN motion in Sect.~\ref{sec:PPK}.} Since the motion takes place in a fixed orbital plane, we do not need to consider the inclination $I$ and the longitude of the node $\Omega$. Thus we have
\begin{subequations}\label{eqRPPR}
\begin{align}
& R = a\left(1-e\cos\psi\right)\,,
\\\
& P^2 = \frac{GM}{a}\,\frac{1+e\cos\psi}{1-e\cos\psi}\,,
\\
& P_R^2 = \frac{GM}{a}\,\frac{e^2\sin^2\psi}{(1-e\cos\psi)^2}\,,
\end{align}
\end{subequations}
together with $\ell=\psi-e \sin\psi$, where $\psi$ is the eccentric anomaly. Then Eq.~\eqref{eqHam} reads
\begin{equation}\label{eqHamaeX}
\frac{H}{\mu} =
-\frac{GM}{2a}
+\frac{1}{2}\left(\frac{GM}{ac}\right)^2\left[
\frac{3\nu-1}{4} + \frac{4-\nu}{\mathcal{X}}-\frac{6+\nu}{\mathcal{X}^2} +\nu\,
\frac{1-e^2}{\mathcal{X}^3}\right]\,,
\end{equation}
where we introduced $\mathcal{X} =1- e\cos\psi$ for convenience. The conjugate
Delaunay-Poincar\'{e} variables $\{\lambda, \Lambda, h, \mathcal{H}\}$ are then
defined by
\begin{subequations}\label{defDelPoincVar}
\begin{align}
&\lambda = \ell +\omega\,, && \Lambda = \mu\sqrt{GMa}\,,
\\
& h = -\omega\,, && \mathcal{H} = \mu\sqrt{GMa}\left(1-\sqrt{1-e^2}\right)\,,
\end{align}
\end{subequations}
and the Hamiltonian equations take the canonical form
\begin{subequations}\label{eqHamDelPoincVar}
\begin{align}
&\frac{\dd\lambda}{\dd t} = \frac{\partial H}{\partial \Lambda}\,, &
\frac{\dd\Lambda}{\dd t} = -\frac{\partial H}{\partial \lambda}\,,\\
& \frac{\dd h}{\dd t} = \frac{\partial H}{\partial\mathcal{H}}\,, &
\frac{\dd\mathcal{H}}{\dd t} = -\frac{\partial H}{\partial h}\,.
\end{align}
\end{subequations}
With the explicit expression of the 1PN Hamiltonian in Eq.~\eqref{eqHamaeX} we obtain
\begin{subequations}\label{eqHamJacobiExplicit}
\begin{align}
&\frac{\dd\Lambda}{\dd t}  =\frac{\dd\mathcal{H}}{\dd t} =
-\frac{\mu\,e}{2\mathcal{X}^3}\left(\frac{GM}{ac}\right)^2\left[\nu-4+2\,\frac{6+\nu}{\mathcal{X}}
-3\nu\,\frac{1-e^2}{\mathcal{X}^2}\right]\sin\psi\,,\\
& 
\frac{\dd \lambda}{\dd t} = n + \frac{GM
n}{2c^2\,a}\left[1-3\nu+\frac{4(\nu-4)}{\mathcal{X}} +
\frac{28+3\nu}{\mathcal{X}^2}-\frac{(4+5\nu)(1-e^2)+12}{\mathcal{X}^3}\right]\nonumber\\
& 
\qquad +\frac{GM n}{c^2(1+\sqrt{1-e^2})a} \left[\frac{\nu-4}{2\mathcal{X}^2} +
\frac{(4+\nu)(1-e^2)+12}{2\mathcal{X}^3} +
\frac{(12+5\nu)(1-e^2)^{3/2}}{2\mathcal{X}^4}-\frac{3\nu(1-e^2)^{5/2}}{2\mathcal{X}^5}\right]\,,\\
& 
\frac{\dd h}{\dd t} = \frac{GM n\sqrt{1-e^2}}{2c^2 e^2\,a\mathcal{X}^2}\left[ 
\nu-4+\frac{(4+\nu)(1-e^2)+12}{\mathcal{X}} -
\frac{(12+5\nu)(1-e^2)}{\mathcal{X}^2}+3\nu\,\frac{(1-e^2)^2}{\mathcal{X}^3}\right]\,.
\end{align}
\end{subequations}

Let us now turn to the integration of the latter equations. As we see from~Eq. \eqref{eqHamJacobiExplicit} we need to obtain the indefinite integrals
$I_n(\psi) = \int \frac{\dd\psi}{\mathcal{X}^n}$ and $J_n(\psi) = \int \dd\psi
\,\frac{\sin\psi}{\mathcal{X}^n}$. Looking at~\citet{GR}, one finds all the needed integrals:
\begin{subequations}\label{eqGR}
\begin{align}
I_1(\psi) &= \frac{2}{\sqrt{1-e^2}}\arctan\left[\sqrt{\frac{1+e}{1-e}}
\tan\frac{\psi}{2}\right]\,,\\
I_n(\psi) &= \frac{e}{(n-1)(1-e^2)}\frac{\sin\psi}{\mathcal{X}^{n-1}} +
(2n-3)I_{n-1}(\psi) - (n-2)I_{n-2}(\psi)\,,\nonumber\\
J_n(\psi) &= - \frac{1}{(n-1)e\mathcal{X}^{n-1}}\,.
\end{align}
\end{subequations}
With these elementary integrals it is straightforward to obtain the complete solution of the motion. We first compute $\Lambda(\psi)$ and $\mathcal{H}(\psi)$. To the zero-th order their values $\Lambda_0$ and $\mathcal{H}_0$ are constant, so the solution is described by constant orbital elements $a_0$, $e_0,$ and $n_0=(G M/a_0^3)^{1/2}$, and we can integrate using $n_0\dd t=\mathcal{X}_0\dd\psi$ to the requested order, where $\mathcal{X}_0=1-e_0\cos\psi$. Using the definitions of $\Lambda$ and $\mathcal{H}$ in~Eq. \eqref{defDelPoincVar} we obtain the solution as $a=a_0+\delta a$ and $e=e_0+\delta e,$ where
\begin{subequations}\label{eq_ae_1PN}
\begin{align}
\delta a &= \frac{GM}{c^2}\left(\frac{1-3\nu}{4}+\frac{\nu-4}{\mathcal{X}_0} +
\frac{6+\nu}{\mathcal{X}_0^2} - \nu\,\frac{1-e_0^2}{\mathcal{X}_0^3}\right) \,,\\
\delta e &= \frac{1-e_0^2}{2a_0\,e_0}\, \delta a\,.
\end{align}
\end{subequations}
In exactly the same way, $\dd h/\dd t$ can be integrated to give $\omega = \omega_0 + \delta\omega$, with
\begin{subequations}\label{eq_omega_1PN}
\begin{align}
\delta\omega = & \, \frac{6\,GM}{a_0 c^2(1-e_0^2)}
\arctan\left[\sqrt{\frac{1+e_0}{1-e_0}}\tan\frac{\psi}{2}\right]\nonumber\\
& +
\frac{GM\sin\psi}{2a_0e_0\sqrt{1-e_0^2}c^2}\left(\frac{(\nu+2)(1-e_0^2)+6e_0^2}{\mathcal{X}_0}+6\,\frac{1-e_0^2}{\mathcal{X}_0^2}-\nu\,\frac{(1-e_0^2)^2}{\mathcal{X}_0^3}\right)\,.
\end{align}
\end{subequations}
Finally, we integrate the mean anomaly $\ell = \psi-e\sin\psi = \lambda + h$. The point is that in the equation for $\lambda$, we first need to express the mean motion as $n=n_0+\delta n,$ where $\delta n=-\frac{3n_0}{2a_0}\delta a$ and $\delta a$ has been found in~Eq. \eqref{eq_ae_1PN}. We thus obtain $\ell=\ell_0+\delta\ell$, where $\ell_0=n_0(t-t_{0,P})=\psi-e_0\sin\psi$ (with $t_{0,P}$ the constant passage at periastron), and
\begin{subequations}\label{eq_l_1PN}
\begin{align}
\delta\ell = & \,\frac{GM(\nu-15)}{8a_0 c^2} \ell_0 \\& -
\frac{GM}{2a_0e_0c^2}\left((4-\nu)e_0^2+\frac{2+\nu+4e_0^2}{\mathcal{X}_0}+6\,\frac{1-e_0^2}{\mathcal{X}_0^2}-\nu\,\frac{(1-e_0^2)^2}{\mathcal{X}_0^3}\right)\sin\psi\,.
\end{align}
\end{subequations}
We note that we could separate, as in Sect.~\ref{sec:PPK}, the secular contributions given by the first terms in both $\delta\omega$ and $\delta\ell$ from the manifestly periodic ones. 

Combining all those equations, we can solve the transit condition to Newtonian order~(Eq. \eqref{cond0}), which gives $\psi_0$, and then the correction $\delta\psi$ coming from Eq.~\eqref{deltaY}, where the modification of the impact parameter is given by perturbing~Eq. \eqref{b0}:
\begin{equation}\label{deltabHam}
\delta b = 
\left\{\begin{array}{llll} \displaystyle b_0\left[\frac{\delta
a}{a_0}-\frac{2e_0\,\delta e}{1-e_0^2} + \frac{\cos\omega_0\,\delta e}{1 -
e_0\cos\omega_0} - \frac{\sin\omega_0\,\delta\omega}{1 - e_0\cos\omega_0}\right]
&&&\text{(transits $T_i$)}\,,\\[0.5cm]
%%%%%%%%%%%%%%%%%%%%%%%%%%%%%%%%%%%%%%%%%%%%%%%%%%%%%
\displaystyle  b_0\left[\frac{\delta a}{a_0}-\frac{2e_0\,\delta
e}{1-e_0^2} - \frac{\cos\omega_0\,\delta e}{1 + e_0\cos\omega_0} +
\frac{\sin\omega_0\,\delta\omega}{1 + e_0\cos\omega_0}\right] &&&\text{(eclipses
$\bar{T}_i$)}\,.
\end{array}\right.
\end{equation}
Finally this gives the modification of the instants of transits due to the 1PN corrections as
\begin{equation}
\delta t = \frac{\mathcal{X}_0 \,\delta\psi- \sin\psi_0\,\delta e- \delta\ell}{n_0}\,,
\end{equation}
where now $\mathcal{X}_0=1-e_0\cos\psi_0$. This method gives the same results as the post-Keplerian ones, reported in Tables~\ref{table1} and~\ref{table2}, up to 0.5\% after 33 cycles.
This difference can be explained by comparing~Eqs. \eqref{deltab} and~\eqref{deltabHam}: the computation of $\delta b$ slightly differs depending on the method used. Moreover by replacing roughly $\varphi_N \rightarrow \varphi_N/(1+k)$ in~Eq. \eqref{deltab}, the second order perturbations can be estimated to play a role at the level of $\sim 5\cdot 10^{-6}$.

\section{Lagrangian perturbation theory}
\label{sec:lag}

It is instructive to work out the problem by means of a Lagrangian perturbation method. In this case we shall focus on secular effects (on a timescale much longer than the orbital period) after averaging the perturbation equations. The Lagrangian of the relative motion of two point masses at the 1PN order is given by
\begin{equation}\label{Lagrangien1PN}
\frac{L}{\mu} = \frac{v^2}{2} - \frac{GM}{r}+ \mathcal{R} \,,
\end{equation}
where the 1PN perturbation function $\mathcal{R}$ depends on the relative position $\bm{x}$ and velocity $\bm{v}=\dd\bm{x}/\dd t$ of the particles (we pose $r=\vert\bm{x}\vert$ and $\dot{r}=\dd r/\dd t$), and reads \citep[see \textit{e.g.}][]{BI03CM},
\begin{equation}\label{Rdef}
\mathcal{R} = \frac{1}{2c^2}\left[\frac{1-3\nu}{4}v^4 + \frac{GM}{r}\left(\nu
\,\dot{r}^2 +(3+\nu)\,v^2\right) - \frac{G^2M^2}{r^2}\right]\,.
\end{equation}
In the Lagrangian perturbation formalism, when looking at secular effects, we can apply the usual perturbation equations of celestial mechanics directly with the perturbation function $\mathcal{R}$ in the Lagrangian \citep[see \textit{e.g.}][]{Dgef94}. As in Sect.~\ref{sec:Ham} we choose the independent orbital elements to be $\{a, e, \ell, \omega\}$, and we pose $n=(G M/a^3)^{1/2}$. The perturbation~(Eq. \eqref{Rexplicite}) depends only on $a$, $e,$ and $\ell$, with the dependence on $\ell$ being implicit in the eccentric anomaly $\psi(e,\ell),$ which is the solution of the Kepler equation, Eq.~\eqref{eqkepler}. The four left perturbation equations read \citep[see \textit{e.g.}][]{BrouwerClemence,BN11}
\begin{subequations}\label{celestmeca}
\begin{align}
\frac{\dd a}{\dd t} &= 
\frac{2}{a n}\, \frac{\partial \mathcal{R}}{\partial \ell}\,,\label{dadt}\\
%%%%%%%%%%%%%%%%%%%%%%%%%%%%%%%%%%%%
\frac{\dd e}{\dd t} &= 
\frac{1-e^2}{e a^2n}\,\frac{\partial \mathcal{R}}{\partial \ell}\,,\label{dedt}\\
%%%%%%%%%%%%%%%%%%%%%%%%%%%%%%%%%%%%
\frac{\dd \ell}{\dd t} &= 
n - \frac{1}{a^2n}\Bigl[2a\,\frac{\partial \mathcal{R}}{\partial a}  +
\frac{1-e^2}{e}\, \frac{\partial \mathcal{R}}{\partial e} \Bigr]\,,\label{delldt}\\
%%%%%%%%%%%%%%%%%%%%%%%%%%%%%%%%%%%%
\frac{\dd \omega}{\dd t} &= \frac{\sqrt{1-e^2}}{e a^2 n}\, \frac{\partial
\mathcal{R}}{\partial e}\,.\label{domegadt}
\end{align}
\end{subequations}
It must be remembered that in perturbation theory $n$ is not equal to $2\pi/P$ with $P$ the period.
Indeed we have the usual definition $\ell=n(t-t_P)$ but the instant of passage at
periastron depends on time: $t_P=t_P(t)$. Instead the period will be given by
averaging  Eq.~\eqref{delldt} for $\ell$. 

As in Sect.~\ref{sec:Ham} we denote the constant orbital elements to zero-th order by
$a_0$, $e_0,$ and pose $n_0=(G M/a_0^3)^{1/2}$. To first order the equations for $a$
and $e$ can readily be integrated as
\begin{equation}\label{integrae}
a = a_0+\frac{2}{a_0 n_0^2}\,\mathcal{R}\,,\qquad e = e_0+\frac{1-e_0^2}{e_0 a_0^2
n_0^2}\,\mathcal{R}\,,
\end{equation}
where the perturbation function reduces in this case to (with
$\mathcal{X}_0=1-e_0\cos\psi$)
\begin{equation}\label{Rexplicite}
\mathcal{R} = \frac{G^2M^2}{2a_0^2c^2}\left[\frac{1-3\nu}{4} +
\frac{\nu-4}{\mathcal{X}_0} + \frac{6+\nu}{\mathcal{X}_0^2}
-\nu\,\frac{1-e_0^2}{\mathcal{X}_0^3} \right]\,.
\end{equation}
The results~of Eqs. \eqref{integrae}--\eqref{Rexplicite} coincide exactly with
those of the Hamiltonian formalism, Eqs.~\eqref{eq_ae_1PN}. Next, the equations for
$\ell$ and $\omega$ to first order become
\begin{subequations}\label{eqsellomega}
\begin{align}
\frac{\dd \ell}{\dd t} &= 
n_0 - \frac{3}{a_0^2 n_0} \mathcal{R} - \frac{1}{a_0^2n_0}\Bigl[2a_0\,\frac{\partial
\mathcal{R}}{\partial a_0}  + \frac{1-e_0^2}{e_0}\, \frac{\partial
\mathcal{R}}{\partial e_0} \Bigr]\,,\\
%%%%%%%%%%%%%%%%%%%%%%%%%%%%%%%%%%%%
\frac{\dd \omega}{\dd t} &= \frac{\sqrt{1-e_0^2}}{e_0 a_0^2 n_0}\, \frac{\partial
\mathcal{R}}{\partial e_0}\,.
\end{align}
\end{subequations}
Finally we have to perform the orbital average of these equations, $\langle f \rangle = \frac{1}{P}\int_0^P\!\! \dd t \,f(t)$. Either we can compute the average directly from Eqs.~\eqref{eqsellomega}, or in a simpler way we can substitute $\mathcal{R}$ by its average $\langle \mathcal{R}\rangle$ in~Eqs. \eqref{eqsellomega}. We have
\begin{equation}\label{avR}
\left\langle \mathcal{R} \right\rangle = \frac{G M a_0 n_0^2}{8
c^2}\biggl(\frac{24}{\sqrt{1-e_0^2}}-15+\nu\biggr)\,,
\end{equation}
hence we end up with the two main results
\begin{subequations}\label{eqaverage}
\begin{align}
\left\langle\frac{\dd \ell}{\dd t}\right\rangle &= n_0 + \frac{GM
n_0\left(\nu-15\right)}{8\,a_0c^2} = n_0\left(1 + \zeta\right)\,,\\
\left\langle\frac{\dd \omega}{\dd t}\right\rangle &= \frac{3G M n_0}{a_0 c^2(1-e_0^2)} = n_0k\,.
\end{align}
\end{subequations}
These are in agreement with the relative modification of the mean motion or period $P$ given by the relativistic parameter $\zeta$ in Eq.~\eqref{zeta}, and of course with the relativistic precession parameter $k$ in Eq.~\eqref{k}. Thus the two averaged Eqs.~\eqref{eqaverage} reproduce the two gauge-invariant equations of the quasi-Keplerian formalism in Sect.~\ref{sec:PPK}. They can also be obtained by averaging the equations of the Hamiltonian formalism in Sect.~\ref{sec:Ham}.

The times of transit and eclipse are straightforwardly computed in perturbation theory using Eqs.~\eqref{eqaverage} and the transit conditions of Eq.~\eqref{geomcondi}. In this way we have confirmed the results reported in Tables~\ref{table1},~\ref{table2}, and~\ref{table3}, but there is a typical difference of nearly $2\%$ after 33 cycles, due to the fact that in this perturbative approach we neglect the periodic terms. As we discussed in Sect.~\ref{sec:PPK}, the relativistic effect on the times of transit after a large number of orbits essentially depends on the precession $k$, and the modification of the orbital period $\zeta$, with the latter effect being smaller (see Eq.~\eqref{ksurzeta}).

\section{Conclusion and discussion}
\label{sec:concl}

In this paper we investigated the relativistic effects in the orbital motion of the high eccentricity exoplanet HD\,80606b orbiting around the G5 star HD\,80606. We proposed a method to detect these effects, based on the accurate measurement of the elapsed time between a mid-transit instant of the planet passing in front of the parent star, and the preceding mid-eclipse instant when it passes behind the star. 

We presented different computations of the relativistic effects on the transit and eclipse times. One is based on the post-Keplerian parametrisation of the orbit in Sect.~\ref{sec:PPK}, another one is a Hamiltonian method with Delaunay-Poincar\'e canonical variables in Sect.~\ref{sec:Ham}, and finally a Lagrangian perturbation approach for computing the secular effects in Sect.~\ref{sec:lag}. A crude understanding and estimation of the main effect is also provided in Appendix~\ref{app:rough}. These methods gave consistent results, which are reported in Tables~\ref{table1},~\ref{table2}, and~\ref{table3}.

We found that in ten to fifteen years, corresponding to 33 to 49 orbital periods, the time difference between the eclipse and the next transit is reduced by approximately 3 to 4.5 minutes due to the relativistic effects for this planet. We conclude that these effects should be detectable for the next observations of the full transit and eclipse of HD\,80606b in coming years by comparing to previous observations done in 2010.

By comparison to short-period exoplanets, only a few long-duration eclipses and transits of HD\,80606b have been observed today. \citet{HD80606b_LDL09} reported the whole eclipse measured with \textit{Spitzer} in November 2007 with a mid-time accuracy of $\pm 260$~seconds. \citet{HD80606b_dW16} secured a similar observation with \textit{Spitzer} of the whole eclipse of January~2010. A few days after again with \textit{Spitzer},~\citet{HD80606b_H11} observed the whole transit of January~2010 with a mid-time accuracy of $\pm 85$~seconds. \citet{HD80606b_R13} observed the same transit with the MOST satellite and reached a lower accuracy of $\pm 294$~seconds. All the other available observations were made from the ground in February~2009 \citep{HD80606b_M09,HD80606b_GM09,HD80606b_F09,HD80606b_Hi10}, June~2009 \citep{HD80606b_W09}, or January~2010 \citep{HD80606b_Sh10} and could only observe partial portions of transits with a given telescope; so their accuracies on the mid-transit times were $\pm 310$~seconds or even poorer.

Thus, the best available measurement of $t_{\rm tr-ec}$ was obtained in January~2010 with \textit{Spitzer} with an accuracy of $\pm\,275$~seconds. A similar accuracy could be reached in 2020 with \textit{Spitzer}, but it would be probably insufficient to detect the effect, slightly below 200~seconds at that time. 
Thanks to its 6.5-m aperture diameter, the \emph{James Webb Space Telescope} (JWST) would provide an accuracy on $t_{\rm tr-ec}$ a few times better than \textit{Spitzer} (0.85-m diameter). So after JWST has started its operation, hopefully in 2021, it should be feasible to detect the effect with its NIRCam or MIRI instruments, while the effect will approach 300 seconds by comparison to 2010. \textit{Spitzer} and JWST are among the rare telescopes able to continuously observe the 12-hour duration transit of HD\,80606b, which is mandatory to reach a good timing accuracy. Other current or future telescopes such as TESS, CHEOPS, or PLATO could also sample long-duration transits; however, they are smaller and less sensitive than \textit{Spitzer} or JWST, so are not able to reach their accuracy on timing. In addition they operate in the optical domain, whereas \textit{Spitzer} and JWST observe in infrared where eclipses are deeper and thus more accurately measured. JWST clearly is the best facility to attempt that detection.\\

Finally we quickly discuss the other perturbing effects that may affect the measurement. HD\,80606b is the only planetary companion detected today in that system. More than 15 years of radial-velocity monitoring\footnote{\citet{HD80606b_HG07} proposed a method to detect non-transiting planetary companions \textit{via} the Newtonian secular precession they induce on an (observed) transiting planet, over and above the effect of general relativity.} did not provide a hint of any additional companion \citep{HD80606b_H11,HD80606b_F16}. With the available data, one can put a $3$-$\sigma$ upper limit of $0.4$\,m/s/yr of a possible linear drift in addition to the radial velocity signature of HD\,80606b; this allows any additional planetary companion with a sky-projected mass larger than $0.5$ Jupiter mass to be excluded with an orbital period shorter than 40~years around HD\,80606. On the other hand, the companion star HD\,80607 is too far away (1200 AU) to produce a noticeable effect on the orbital precession of HD\,80606b. 

Besides the possibility that there might be some disturbing other bodies, we have to consider the precession of the orbit caused by the oblateness of the star, the tidal effects between the star and the planet, and the Lense-Thirring effect. The orbital precession rate due to the quadrupolar deformation $J_2$ of the star is
\begin{equation}\label{DeltaJ2}
\Delta_{J_2} = \frac{3\pi J_2\,R_\star^2}{a^2 (1-e^2)^2}\,.
\end{equation} 
Assuming the solar value $J_2\sim 10^{-7}$ we obtain $\Delta_{J_2} \sim 0.4\,\text{arcsec}/\text{century}$ for HD\,80606b, which is negligible for our purpose. As for the orbital precession induced by the tidal interaction of the planet with its parent star, it produces a supplementary orbital precession at the rate \citep[see \textit{e.g.}][]{HD80606b_JB09,HD80606b_F10}
\begin{equation}\label{DeltaT}
\Delta_\text{T} = 30 \pi \left( k_p\,\frac{M_\star R_p^5}{M_p} + k_\star\,\frac{M_p
R_\star^5}{M_\star}\right)\frac{1+\frac{3}{2}e^2+\frac{1}{8}e^4}{a^5(1-e^2)^5}\,,
\end{equation} 
where $k_p$ denotes the Love number of the planet, which is approximately $k_p\sim 0.25$ for a hot Jupiter, and $k_\star$ that of the star, expected to be approximately $k_\star\sim 0.01$~\citep{HD80606b_JB09}. We find that the first term, due to the tidal deformation of the planet, gives a rather large contribution of about $32\,\text{arcsec}/\text{century}$, while the second term in Eq.~\eqref{DeltaT}, due to the deformation of the star, is much smaller, approximately $2\,\text{arcsec}/\text{century}$. Hence the tidal interaction represents a non-negligible effect, but nevertheless one that is smaller than the GR relativistic precession:
\begin{equation}\label{DeltaTGR}
\Delta_\text{T} \sim 34 \,\text{arcsec}/\text{century} \simeq 0.16
\,\Delta_\text{GR}\,.
\end{equation} 
Hence we conclude that the tidal interaction should be taken into account in the precise data analysis of the transit times aimed at measuring the GR effect.
The  Lense-Thirring effect, due to the angular momentum $S_\star$  of the star, induces a precession of the line of nodes with a rate of
\begin{equation}\label{LenseThirring}
\frac{\dd \Omega}{\dd t} = \frac{2G\, S_\star}{c^2a^3(1-e^2)^{3/2}}\,,
\end{equation} 
when $M_p\ll M_\star$. Considering the extreme case of a spin-orbit alignment and assuming a solar value $S_\star\sim 10^{42} \,\text{kg.m}^2/\text{s}$, it is negligible, as expected: $\Delta_\text{LT} \sim 0.07\,$arcsec/century.

To compare with the solar system, those three effects are negligible for Mercury, for which they are on the order of $\Delta_{J_2} \simeq 0.2\,$arcsec/century, $\Delta_\text{T} \simeq 2\cdot 10^{-6}\,$arcsec/century, and $\Delta_\text{LT} \sim 2\cdot 10^{-3}\,$arcsec/century. In the case of Mercury the periastron advance is largely dominated by the Newtonian effect of the other planets, mainly Venus and Jupiter: $\Delta_\text{planets} = 532.3\,$arcsec/century $\simeq 12.4 \,\Delta_\text{GR}$. From this point of view, HD\,80606/HD\,80606b represents a cleaner system than the solar system.

\begin{acknowledgements}
The authors would like to thank Alain Lecavelier des \'Etangs and Gilles Esposito-Far\`ese for motivating discussions at an early stage of the work.
\end{acknowledgements}

\begin{appendix}

\section{Rough estimation of the effect}
\label{app:rough}

\begin{figure}[h]
\centering
        \includegraphics[scale=.6]{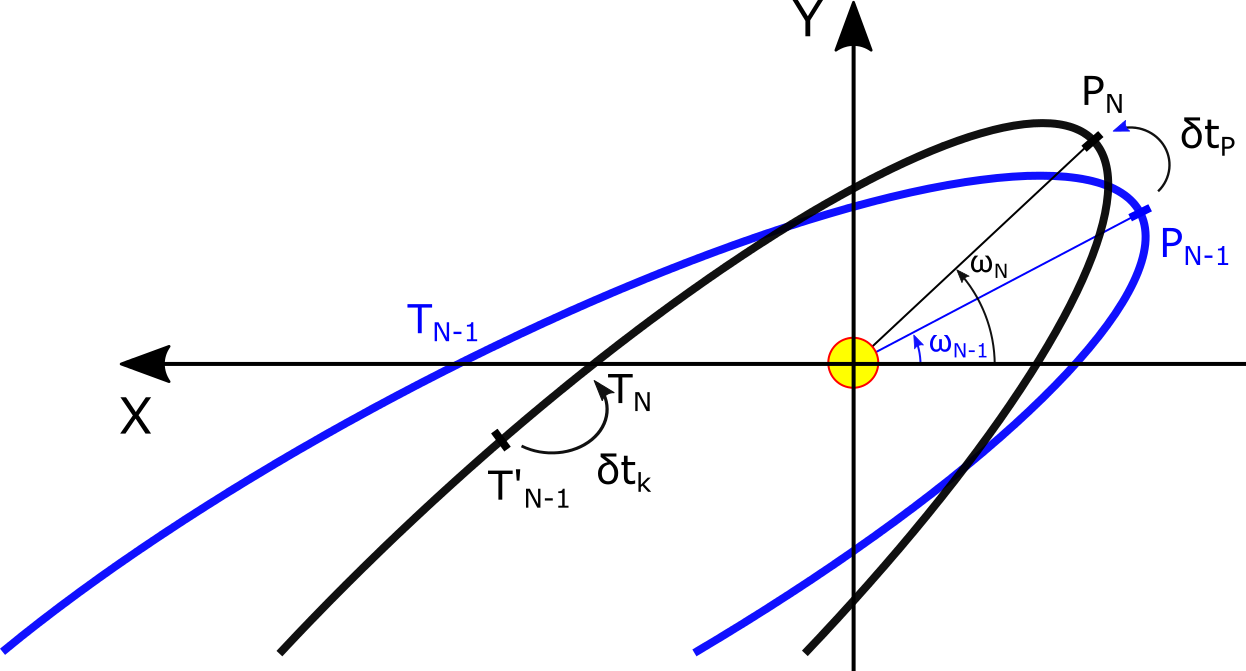}
        \caption{Displacement of the trajectory and time shifts of the exoplanet in-between two approximate successive Keplerian orbits $N-1$ and $N$. The two main contributions $\delta t_k$ and $\delta t_P$ of the (rough) estimation of the time shift are indicated. The point $T'_{N-1}$ is the point corresponding to $T_{N-1}$ reported in the next orbit $N$.}
        \label{fig_dt_gross}
\end{figure}

It is also possible to estimate the PN effects by considering a step-by-step shift of the orbit along an approximate sequence of Keplerian ellipses. As depicted in Fig.~\ref{fig_dt_gross}, at each step we have to consider two time delays: the one induced by the new position of the transit on the orbit, $\delta t_k$, and the time to pass from the previous periastron to the current one, $\delta t_P$.

The first delay is due to the different orientation of the $N$-th Keplerian ellipse with respect to the $(N-1)$-th one, shifted by the relativistic precession angle $\omega_N-\omega_{N-1} \simeq 2\pi k$ (see Fig.~\ref{fig_dt_gross}). It simply reads $n_0\,\delta t_k \simeq \ell_{N} - \ell_{N-1}$, where $n_0$ is the Newtonian mean motion and $\ell_{N}$ the mean anomaly solving the transit condition of the $N$-th orbit (for a given transit $T_i$ or eclipse $\bar{T}_i$). Of course the mean anomaly is counted from the periastron $P_N$ of the $N$-th orbit, and is given in terms of the eccentric anomaly by Kepler's equation $\ell_N=\psi_N-e \sin\psi_N$. The condition relating the corresponding true anomalies of the successive ellipses is thus
\begin{equation}
\arctan \biggl(\sqrt{\frac{1+e}{1-e}} \tan \frac{\psi_N}{2} \biggr) - \arctan
\biggl(\sqrt{\frac{1+e}{1-e}} \tan \frac{\psi_{N-1}}{2} \biggr) + \pi k \simeq 0\,.
\end{equation}

As for the second time delay, it is due to the modification of the mean motion (and thus the period) as given by $n=n_0(1+\zeta)$ (see Eq.~\eqref{zeta}). We have $\ell = n_0(t-t_{P_N})$, which we can roughly equate to $n(t-t_{P_{N-1}})$, where $t_{P_{N-1}}$ and $t_{P_N}$ are the successive instants of passage at periastron. Thus $t_{P_N}-t_{P_{N-1}}\simeq-\zeta(t-t_{P_{N-1}})$, and we roughly approximate $t-t_{P_{N-1}}$ by the period so that $n_0\,\delta t_P \simeq - 2\pi \zeta$. The time delay between the $N$-th orbit and the reference one finally becomes
\begin{equation}
\delta t(N) \simeq \frac{\psi_N - \psi_0 - e(\sin\psi_N - \sin\psi_0) - 2\pi N \zeta}{n_0}\,.
\end{equation}
It is clear from Fig.~\ref{fig_dt_gross} that the two effects are negative, and add up negatively in the total delay. Even if this method seems quite crude, we find that it gives a reasonable estimate of the time shifts. Indeed, the computed times differ from the ``exact'' post-Keplerian results reported in Tables~\ref{table1}--\ref{table3} by only 9\% after 33 cycles.

\end{appendix}

\bibliographystyle{aa}
\bibliography{ListeRef_HD80606b}

\end{document}